# Horizon-T Experiment Upgrade and Calibration of New Detection Points

D. Beznosko [1a], A. Iakovlev [a], S. Jakupov [a], D. Turganov [a], A. Tussipzhan [a], T. Uakhitov [a], M.I. Vildanova [b], A. Yeltokov [a], V.V. Zhukov [b]

*Abstract*

In March of 2018, after the completion of the Physics Run 2, an upgrade has been installed at an innovative detector system Horizon-T, with the upgraded version now called Horizon-10T. It was constructed to study Extensive Air Showers (EAS) in the energy range above $10^{16}$ eV coming from a wide range of zenith angles (0º - 85º). The system is located at Tien Shan high-altitude Science Station of Lebedev Physical Institute of the Russian Academy of Sciences at approximately 3340 meters above the sea level.

After this upgrade, the detector consists of ten charged particle detection points separated by the distance up to 1.3 kilometer as well as optical detector to view the Vavilov-Cherenkov light from the EAS. Each detector connects to the Data Acquisition system via cables. The calibration of the time delay for each cable including newly installed ones and the signal attenuation is provided in this article as well as the description of the newly installed detection points and their MIP response values.

## 1. Detector System Description

"Horizon-T" detector system [1] [2] [3] [4] is used to study EAS with parent particle of energies higher than $10^{16}$ eV coming from a range of zenith angels (0º-85º). The naming convention is Horizon-xT, where x is the number of detection points functioning at the start of physics run with that detector version. Current upgrade has brought the number of detection points to 10, thus making the detector system name of Horizon-10T [24], [25].

"Horizon-T" is constructed to study space-time distribution of the charged particles in EAS disk and Vavilov-Cherenkov radiation from it. The novel method of using time information from pulse shape in each detector allows for the analysis of EAS with core falling outside of the detector system bounds. It is located at the Tien Shan High-Altitude Science Station (TSHASS), a branch of the Lebedev Physical Institute of the Russian Academy of Science. It is located 32 km from Almaty at the altitude of 3340 meters above the sea level.

Time of passage of the charged particles from EAS disk are registered at ten detection points. The relative coordinates of every point are presented in the Table 1. The Horizon-10T detector system aerial view is presented in Figure 1. The system center is marked by a geodesic benchmark installed at the detection point 1 at the height of 3346.05 meters above the sea level and with geographical coordinates of 43°02′49.1532" N and 76°56′43.548" E. This benchmark is the origin for the XYZ coordinate system for Horizon-10T. The X-axis is directed to the north, Y-axis to the west and Z-axis is directed vertically up. The geometric factor of the detector system is ~1.5 km²/ster at $10^{17}$ eV. The detector of Vavilov-Cherenkov radiation is located next to station 1. The signal-carrying cables of all three Cherenkov detectors are of the same length. Work is in progress to build a cable-less version of the detector system called HT-KZ [5] [6].

---
[1] dmitriy.beznosko@nu.edu.kz (also dima@dozory.us)
[a] Physics Department, Nazarbayev University, Astana, Kazakhstan
[b] P. N. Lebedev Physical Institute of the Russian Academy of Sciences, Moscow, Russia

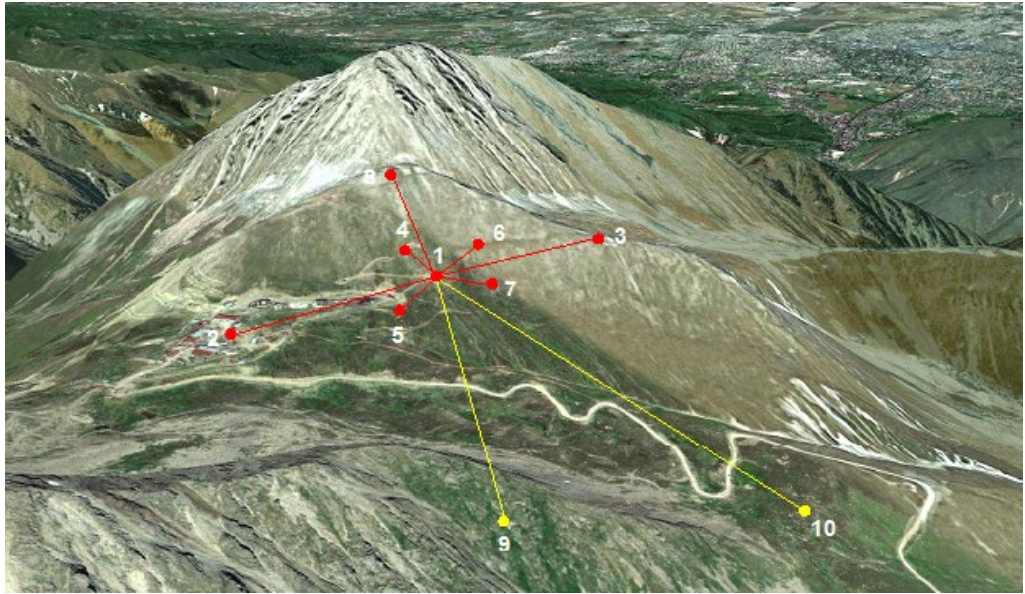

Figure 1: Detection points aerial view.

Table 1: Coordinates of all detection points.

| Station # | X, m | Y, m | Z, m | R, m |
|---|---|---|---|---|
| 1 | 0 | 0 | 0 | 0 |
| 2 | –445.9 | –85.6 | 2.8 | 454.1 |
| 3 | 384.9 | 79.5 | 36.1 | 394.7 |
| 4 | –55.0 | –94.0 | 31.1 | 113.3 |
| 5 | –142.4 | 36.9 | –12.6 | 147.6 |
| 6 | 151.2 | –17.9 | 31.3 | 155.4 |
| 7 | 88.6 | 178.4 | –39.0 | 203.0 |
| 8 | 221.3 | 262.0 | 160.7 | 378.7 |
| 9 | | | | 600 |
| 10 | | | | 1000 |

## 1.1 Charged Particles Detectors and Data Acquisition System

Currently detection points 1 through 8 have just one scintillator detector (SD) which is in the z-plane (e.g. it is parallel to the sky). Previously located in these points detectors in x and y planes have been moved to detection points 2, 9 and 10. Due to the weather conditions, a previously planned upgrade to liquid scintillators [7] [8] is not being considered.

Each SD in points 1-8 uses polystyrene-based square-shaped cast scintillator [9] with 1 m$^2$ area and 5 cm width. 2-inch Hamamatsu [10] R7723 photoelectrical multipliers (PMT) register scintillator light. Additionally, points 1 and 4-7 have fast time resolution glass-based detectors (GD) [11] with R7723 PMT as a readout.

All PMT signals are carried without any amplifiers over the coaxial cables RK 75-7-316F-C SUPER produced by SpetsKabel [12] and impedance matched to the rest of the electronics and calibrated [13]. The system-wide electronic trigger is formed by a first 14-bit CAEN [14] DT5730 ADC (analog to digital converter) board. Three ADC boards (same model) in a common trigger schema make the data acquisition system (DAQ) system located immediately next to the detection point #1. The DAQ is triggered when the detection points 4 and 7 report the passage of charged particles from EAS disk. This initial hardware trigger allows keeping a larger data sample for further offline analysis. Typical offline trigger requires a signal from all four detection points (4, 5, 6 & 7).

Detection points 1 - 3 have PMT-49 (FEU49) from MELZ [15] (a 15cm diameter spherical-shaped cathode PMT with the spectral response from 360nm to 600nm) installed alongside the R7723 PMT so cross-calibration between these PMTs is possible. They are also used for the MIP calibration in a double coincidence triggering schema The important parameter to know is the pulse time difference between the models, which is determined by the electrons time of flight within the PMT, and it is ~48ns with a 450m cable.

During the upgrade of 2018, the detection point 2 now has additional 5 SDs with FEU49 in addition to the previously existing SD with R7723 PMT. The outputs of these five detectors is summed and is transmitted over the single cable. Detection points 9 and 10 use a SD with 12.7 cm Hamamatsu H6527 PMTs. In addition, point 9 has 4 SDs with FEU49 with the output summed and detection point 10 has more than that – 11 SDs with FEU49 with the output summed, but these SD with older PMTs are not connected to DAQ at this time. The need for this upgrade was shown from previous physics run results [16] that signal from EAS at large distances from the axis pose most interest but with lower particle density a higher detection area is needed.

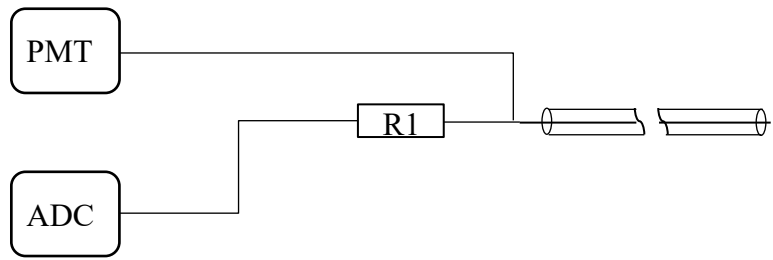

Figure 2: Schematic of the cable calibration setup

## 2. Calibration of Cables and MIP Response

Previous cable calibrations are described in [13]. The same procedure was used (briefly re-capped below).

All PMTs are connected to the DAQ system via coaxial cables of length ranges from 20m to ~1000m. The common setup used is shown in Figure 2. The cable is disconnected at one end to obtain pulse reflection from it. R1 is the 25Ω resistor to achieve the impedance matching between the 50Ω for the ADC, and the cable that has impedance of 75Ω.

A stand-alone PMT49 with a 15cm x 15cm x 1cm plastic scintillator piece provides a test signal from cosmic muons. The pulses are recorded by the 14bit CAEN [14] DT5730 flash ADC. The PMT bias is -1500V and 40mV threshold was set to select cosmic ray events. Events are then analyzed.

The MIP calibration procedure is detailed in [17]. Each SD/GD response to MIP is calibrated individually. For triggering, a detector consisting of FEU49 with a 15 cm diameter scintillator is placed under a detector under calibration. Double-coincidence schema is used, facilitated by the ADC. The setup schematic is shown in Figure 3. Since only two cables connect each detection point with DAQ physically, thus only a double coincidence setup was realized.

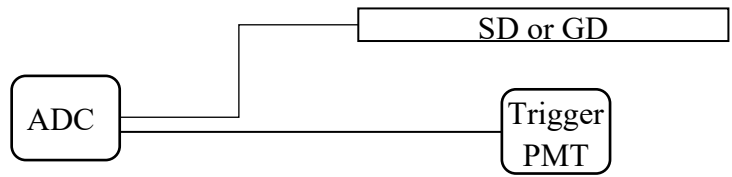

**Figure 3: MIP calibration setup schematic for SD/GD.**

The calibration process yields the area of a single MIP signal as well as the width. The total pulse duration is taken as time between the 0.1 and 0.9 of the pulse area; pulse front is defined as time between 0.1 and 0.5 of the pulse area. This choice reduces the baseline noise effects. The uncertainty, associated with the size of the integration window is included in the total error.

## 2.1     Data Analysis Results

Detection points are labeled with numbers, but also have their 'internal' names as well as codes for the cables. The calibration data is given using the internal names for the convenience of this document use. Table 2 gives detection point numbers and their internal names.

**Table 2: Names of detection stations**

| *Station name* | Center | Yastrebov | Stone Flower | Left | Kurashkin | Right | Bottom | Upper | Cher | 600m | Bunker |
|---|---|---|---|---|---|---|---|---|---|---|---|
| *Station number* | 1 | 2 | 3 | 4 | 5 | 6 | 7 | 8 | VCD | 9 | 10 |

The time calibration values and the ratios of the areas obtained from the ADC data directly are given in the Table 3. The pulse widening is noticeable from the simple comparison of the time differences at 10, 50 and 90 percent of the pulse area. Specifically, this is very pronounced for the farthest stations, such as 2, 3 and 8-10.

**Table 3: Time differences and area ratios from ADC data**

| Station and cable | Time diff. at 10% (ns) | Time diff. at 50% (ns) | Time diff. at 90% (ns) | Area ratio |
|---|---|---|---|---|
| Bottom New | 763.6 ± 0.106 | 767.1 ± 0.226 | 784.9 ± 1.688 | 0.548 ± 0.00797 |
| Bottom Old | 1016.54 ± 0.135 | 1021.26 ± 0.779 | 1043.162 ± 7.109 | 0.723 ± 0.0544 |
| Center Blue | 72.987 ± 0.0491 | 73.552 ± 0.063 | 75.809 ± 0.465 | 0.6711 ± 0.00875 |
| Center Blue New | 102.294 ± 0.0771 | 103.476 ± 0.0956 | 109.541 ± 0.82 | 0.727 ± 0.00759 |
| Center Green | 73.14 ± 0.08 | 74.176 ± 0.06 | 76.56 ± 0.36 | 0.598 ± 0.003 |
| Center Red | 72.864 ± 0.076 | 73.383 ± 0.0695 | 75.19 ± 0.449 | 0.67 ± 0.00526 |
| Center Red New | 102.35 ± 0.08 | 103.49 ± 0.08 | 108.83 ± 0.50 | 0.72 ± 0.005 |
| Center White | 73.508 ± 0.0775 | 74.045 ± 0.0746 | 76.177 ± 0.446 | 0.669 ± 0.00583 |
| Center Yellow | 72.979 ± 0.06198 | 73.472 ± 0.0771 | 75.656 ± 0.347 | 0.669 ± 0.00506 |
| Cher Green Red | 88.12 ± 0.09 | 88.91 ± 0.06 | 92.63 ± 0.43 | 0.583 ± 0.004 |
| Cher White Blue | 82.18 ± 0.12 | 83.25 ± 0.07 | 87.47 ± 0.41 | 0.583 ± 0.004 |
| Cher Yellow | 81.42 ± 0.11 | 82.46 ± 0.08 | 86.44 ± 0.62 | 0.582 ± 0.003 |
| Kurashkin New | 709.0448 ± 0.109 | 711.77 ± 0.235 | 723.762 ± 1.815 | 0.583 ± 0.0131 |
| Kurashkin Old | 881.246 ± 0.102 | 883.467 ± 0.251 | 893.162 ± 2.513 | 0.521 ± 0.0178 |
| Left New | 551.372 ± 0.091 | 553.464 ± 0.1797 | 562.726 ± 1.275 | 0.596 ± 0.0101 |
| Left Old | 1058.303 ± 0.0924 | 1061.044 ± 0.226 | 1073.422 ± 2.205 | 0.585 ± 0.0158 |
| Right New | 561.744 ± 0.086 | 564.00 ± 0.127 | 574.68 ± 1.157 | 0.669 ± 0.00506 |
| Right Old | 714.608 ± 0.202 | 716.493 ± 0.504 | 720.142 ± 2.988 | 0.3531 ± 0.0724 |
| Right Old2 | 770.374 ± 0.116 | 773.482 ± 0.184 | 788.544 ± 1.16 | 0.746 ± 0.096 |
| Stone Flower New | 1601.95 ± 0.148 | 1609.835 ± 1.963 | 1645.642 ± 13.724 | 0.501 ± 0.0707 |
| Stone Flower Old | 2329.274 ± 0.15 | 2335.899 ± 0.5998 | 2365.196 ± 3.867 | 0.521 ± 0.0189 |
| Yastrebov New | 2132.105 ± 0.188 | 2143.136 ± 2.794 | 2189.78 ± 22.554 | 0.461 ± 0.0961 |
| Yastrebov Old | 2553.235 ± 0.161 | 2561.865 ± 2.937 | 2599.027 ± 19.102 | 0.528 ± 0.128 |
| Upper New using function generator | 2049.13 ± 0.05 | 2057.35 ± 0.05 | 2094.28 ± 0.21 | 0.337 ± 0.0008 |
| Upper New | 2050.728 ± 0.183 | 2061.111 ± 1.464 | 2105.033 ± 9.093 | 0.474 ± 0.0461 |
| 600m | 2836.205±0.212 | 2855.306±1.925 | 2925.0115±7.377 | 0.408±0.0288 |
| Bunker | 4583±0.854 | 4638.342±17.186 | 4795.407±42.232 | 0.3755±0.1328 |

The results for the MIP tests are only presented for three detection points, but the table will be filled as data analysis will complete. The current results are presented in Table 4. All effects from cables are accounted for in the calibration; the shown values are: MPV (top value) and σ (bottom value) with corresponding fit uncertainties.

**Table 4: Detector MIP response pulse area at operating bias voltage**

| Detection point and cable designation | Detector type | Area (ADC counts · ns) | Detection point and cable designation | Detector type | Area (ADC counts · ns) |
|---|---|---|---|---|---|
| 600m | SC | 5782 ± 198<br>1445 ± 108 | Bunker | SC | 6583 ± 107<br>984 ± 76 |
| Upper New | SC (Since March 1, 2017, before March 22, 2018) | 5713 ± 97<br>640 ± 56 | Center Yellow | SC (after March 20, 2018) | 215 ± 6<br>42 ± 4 |
| Upper New | SC (after March 22, 2018, 2pm) | 1118 ± 18<br>150 ± 11 | Center White | SC (FEU49) | 1125 ± 25<br>214 ± 16 |

## 3. Conclusion

The upgrade and new cables calibration of Horizon-T has been completed. The time delay of the pulse at three different area fraction points is done for each pulse, thus showing the pulse widening as it travels along the cable. Widening is different between PMT and generator pulses so PMT is used for all channels. The losses in each cable are also monitored using the ratio of the areas of the reflected pulse over the original, taken between the same fractions of each pulse as the timings. This also indicates that any further calibrations of the SD and VCD should be done via their cables to include any effects implicitly, and this is used for the MIP calibration of each detection point.

## Acknowledgement

This work is funded in part by MES RK state-targeted program BR05236454.